\documentclass[%
 reprint,
 amsmath,amssymb,
 aps,
]{revtex4-1}

\usepackage{amsmath}
\usepackage{amssymb}
\usepackage[usenames, dvipsnames]{color}
\usepackage{graphicx}
\usepackage{dcolumn}
\usepackage{bm}

\usepackage{listings}
\usepackage{color}

\definecolor{dkgreen}{rgb}{0,0.6,0}
\definecolor{gray}{rgb}{0.5,0.5,0.5}
\definecolor{mauve}{rgb}{0.58,0,0.82}

\begin{document}

\preprint{APS/123-QED}

\title{Nuclear Physics without High-Momentum Potentials: \\
Direct Construction of the Effective Interaction from Scattering Observables}

\author{Kenneth S. McElvain}
 \email{kenmcelvain@berkeley.edu}
  \author{W. C. Haxton}
\email{haxton@berkeley.edu}
\affiliation{Department of Physics,  University of California, Berkeley, and Lawrence Berkeley National Laboratory, Berkeley CA
}


\date{\today}

\begin{abstract}
The standard approach to nuclear physics encodes phase shift information in an NN potential,
then decodes that information in forming an effective interaction, appropriate to a low-momentum Hilbert space.  Here we show that
it is instead possible to construct the effective interaction directly from 
continuum phase shifts and mixing angles, 
eliminating all reference to a high momentum potential.  The theory is rapidly convergent and well behaved, yielding sub-keV accuracy.

\end{abstract}

\pacs{Valid PACS appear here}
\maketitle



%
  
Traditional nuclear physics is based on an encoding of experimental phase shift information into an
NN potential, followed by renormalization to obtain an effective interaction appropriate for soft, discrete bases,
such as those used in the shell model. 
This approach has proven problematic, due to the strength 
of the bare interaction, its extreme hard core, and its disparate length scales.  Diagrammatic effective
interaction methods 
were found to fail in the early 70s \cite{BK,SW}; in recent years some difficulties have been ameliorated, with novel techniques introduced to soften interactions, and with computing power allowing use of much larger effective Hilbert spaces
\cite{GW,Bogner,vlowk,nocore,renorm}.  Yet aspects of these techniques remain approximate.

One can ask why this approach is taken.  The nuclear physics ``two-step" -- from QCD to an NN
potential to an $H^\mathrm{eff}$ appropriate for some discrete Slater determinant basis -- differs from 
standard effective field theory methods, where the reduction from the fundamental ultraviolet (UV) theory
to the effective infrared (IR) theory is generally direct.  There is no obvious reason why,
in nuclear physics, it is necessary to store UV information in an
NN potential, if in the end all UV details are integrated out, in the process of finding
effective interactions appropriate for restricted Hilbert spaces.

The effective theory (ET) employed in any direct construction must have certain properties.
The functional form of $H^\mathrm{eff}$  must be known, before its parameters can be fit.
The theory should be analytically continuous in $E$ --  valid for $E<0$ and $E>0$ -- if scattering data
is to be used in the fit, with bound-state properties then predicted.  Translational invariance is critical to ensuring a
simple functional form for $H^\mathrm{eff}$.
If one formulates an ET in a discrete, compact basis, suitable for the powerful diagonalization methods 
that have been developed in nuclear physics, this limits the choice to center-of-mass (CM) separable bases of the 
harmonic oscillator (HO).

Continuity in $E$ is generally not a feature of the approximate effective interactions used in nuclear
physics.  On the contrary, work has been invested to remove any energy dependence from
effective interactions, through techniques such as
the Lee-Suzuki transformation \cite{LS}.  Consequently properties one expects in a well-defined effective theory,
such as effective wave functions that correspond to the projections of the true wave functions, are absent:
projection does not preserve orthogonality, while the Hermitian, energy-independent interactions in common
use clearly do.  This leads to an odd contrast between bound state treatments and scattering,
troublesome from the perspective of energy continuity, 
as much of the information extracted from phase shifts is
associated with the unusually rapid variations with energy caused by anomalously large scattering lengths.

\begin{figure}[ht]   
\centering
\includegraphics[scale=0.27]{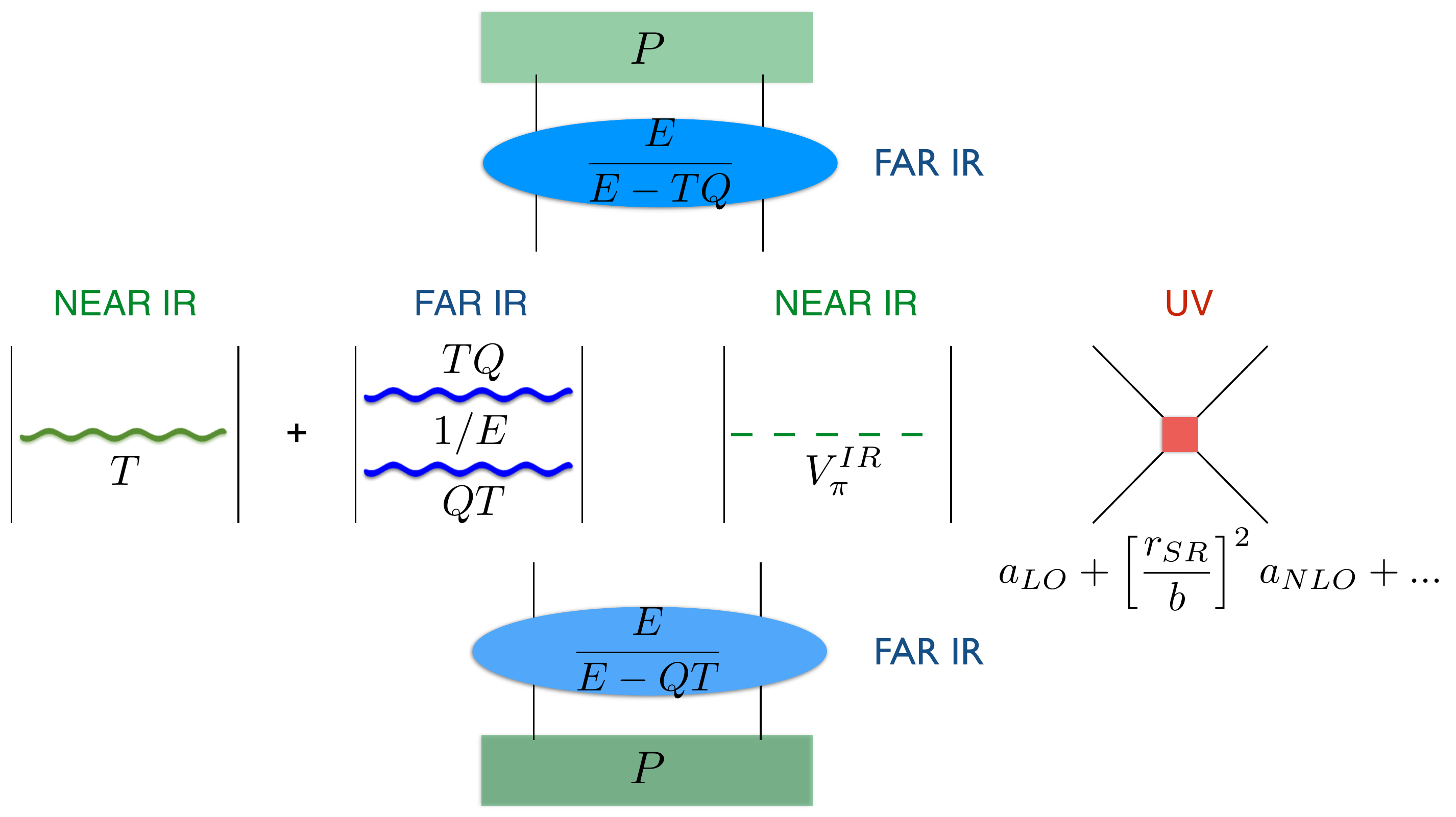}
\caption{HOBET's pionful effective interaction, appropriate to a HO where translational invariance 
requires $P$ to be defined in terms of total quanta (in contrast to chiral interactions employing a momentum cut). (Color online: blue, green, red
indicate far-IR, near-IR, and UV corrections.)}
\label{Fig_Heff}
\end{figure}

A nonrelativistic ET for simple nuclear systems -- HO-based effective theory (HOBET) --
was constructed several years ago and applied to bound states \cite{WH}.  The functional form of the two-body HOBET effective interaction
was deduced from exact numerical solutions, and found to correspond to a
rapidly converging short-range effective theory -- but only if the underlying Bloch-Horowitz (BH) equation \cite{BH}
is rearranged in the following way,
\begin{eqnarray}
&&~~~~~~~~~~~~~~~~~~~P H^\mathrm{eff} P | \Psi \rangle = E P | \Psi \rangle \nonumber \\
&&G_{QT} \equiv  {1 \over E - QT}~~~~~G_{QH} \equiv {1 \over E-QH}~~~~~~H \equiv T+V \nonumber \\
&&~~H^\mathrm{eff} = E G_{TQ}(E)  \left[ T + T {Q \over E} T + V + V_\delta \right] E G_{QT}(E)  \nonumber \\
&&~~~~~~~~~~~~~~~~~~~~~~V G_{QH} QV  \leftrightarrow  V_\delta 
\label{eq:BH}
\end{eqnarray}
(See Fig. \ref{Fig_Heff}.) This reordering respects an important condition on ETs based on short-range expansions, that they can succeed only if the
proper IR behavior is built in \cite{Lepage}: We recognize $E G_{QT}(E)$ as
the asymptotic Lee-Suzuki operator, which generates the full IR solution from the projected wave function.  
Here $P=P(b,\Lambda)$ is the separable projected space, defined by the oscillator
parameter $b$ and the maximum number of oscillator quanta $\Lambda$ allowed in Slater determinants,
and $Q$ is its complement.  The equation must be solved self-consistently, a step that determines
bound-state eigenvalues.  Regardless of the dimensionality of $P$, the BH equation generates all
solutions having nonzero overlaps with $P$, yielding exact eigenvalues and projected
wave functions $P|\Psi \rangle$.

The choices made for $V_\delta$ and $V$ in Eq. (\ref{eq:BH}) define the ET.  In HOBET's original validation,
$V$ was equated to the Argonne $v18$ potential \cite{av18}; the associated
scattering in $Q$ from the fourth line of Eq. (\ref{eq:BH}) was computed numerically \cite{num};  and the results were shown to be equivalent, to
very high accuracy, to the short range expansion represented by $V_\delta$ \cite{WH}.  This is the traditional renormalization procedure,
though with the end result expressed compactly in terms of HOBET's effective interaction.

Here we execute HOBET as a true ET, severing connections to realistic potentials.
Instead of numerically integrating out a UV potential, the parameters of the ET are determined directly
from the energy self-consistency of the BH equation, after imposing appropriate IR boundary conditions through $G_{QT}$.
Bound state wave functions must vanish asymptotically, while
scattering states must oscillate with the proper phase shift \cite{lattice}.
In analogy with nuclear EFT, two versions are natural, pionless HOBET where $V \equiv 0$,
and pionful HOBET where $V \rightarrow V_\pi^{IR}$.   
$V_\pi^{IR}$ is a further IR correction of long-distance behavior, augmenting the kinetic energy
summation.  In contrast to EFT approaches, the pion plays no role at short ranges in HOBET.

As HOBET's technical aspects are described elsewhere \cite{WK}, the description here focuses on simple examples.
The original HOBET expansion \cite{WH}
can be recast in terms of HO
creation $(a^\dagger_x,a^\dagger_y,a^\dagger_z)$  and annihilation operators, 
\[ a_i\equiv{1 \over \sqrt{2}} \left( {\partial \over \partial r_i} + r_i \right)~~~~~~a_i^\dagger \equiv {1 \over \sqrt{2}} \left( -{\partial \over \partial r_i} + r_i \right) \]
satisfying the usual commutation relations.
Here $\boldsymbol{r}=(\boldsymbol{r}_1-\boldsymbol{r}_2)/\sqrt{2} b$ is the dimensionless Jacobi coordinate.
Defining projections with good angular momentum,
 $a^\dagger_M = \hat{\boldsymbol{e}}_M \cdot \boldsymbol{a}^\dagger$ and $\tilde{a}_M=(-1)^M a_{-M}$, where
 $\hat{\boldsymbol{e}}_M$ is the spherical unit vector, nodal and angular momentum
 lowering operators for the HO can be formed
 \begin{eqnarray}
  && {\bf \tilde{a}} \odot {\bf \tilde{a}}~ | n \ell m \rangle = -2\; \sqrt{\left(n-1\right)\left(n+\ell-{1/ 2}\right) }  \; |n-1 \, \ell m \rangle \nonumber \\
  && \left[ \left[ {\bf \tilde{a}} \otimes {\bf \tilde{a}} \otimes \cdots \otimes {\bf \tilde{a}}\right]_\ell \otimes |n \ell \rangle \right]_{00} = \nonumber \\
 &&~~~~~~~~~~~~~   (-1)^\ell \, 2^{\ell/2}  \sqrt{{l! \over (2 \ell-1)!!} {\Gamma[n+\ell+{1 \over 2}] \over \Gamma[n+{1 \over 2}]}} ~|n 0 0 \rangle 
 \end{eqnarray}
where $|n \ell m \rangle$ is a normalized HO state.  Using
\begin{eqnarray}
 \delta(\boldsymbol{r}) &=& \sum_{n^\prime n} d_{n^\prime n}^{00} |n^\prime 0 0 \rangle \langle n 00 | \nonumber \\
 d_{n^\prime n}^{\ell^\prime \ell} &\equiv& {2 \over \pi^2} \left[ {\Gamma(n^\prime+\ell^\prime +{1 \over 2}) \Gamma(n+\ell+{1 \over 2}) \over (n^\prime-1)! \, (n-1)!} \right]^{1/2} ,
 \end{eqnarray}
 HOBET's short-range expansion can be carried out.  We obtain the S channel N$^3$LO and abbreviated SD(tensor interaction) channel expansions
 \begin{widetext}
  \begin{eqnarray}
&& V_\delta^{\mathrm{S}} =\sum_{n^\prime n} d_{n^\prime n}^{\, 0 \, 0} \, \Big[ a^{\mathrm{S}}_{LO} \, | n^\prime \, 0 \rangle \langle n \, 0 |  + a_{NLO}^{\mathrm{S}} \, \left\{ {\bf a}^\dagger \odot {\bf a}^\dagger  | n^\prime \, 0 \rangle \langle n \, 0 | +  | n^\prime \, 0 \rangle \langle n \, 0 | {\bf \tilde{a}} \odot {\bf \tilde{a}} \right\} + a_{NNLO}^{\mathrm{S} , 22} \, {\bf a}^\dagger \odot {\bf a}^\dagger  | n^\prime \, 0 \rangle \langle n \, 0 |{\bf \tilde{a}} \odot {\bf \tilde{a}} + \nonumber \\
 &&~
 a_{NNLO}^{\mathrm{S} ,40} \, \left\{ ({\bf a}^\dagger \odot {\bf a}^\dagger )^2 | n^\prime \, 0 \rangle \langle n \, 0 | +  | n^\prime \, 0 \rangle \langle n \, 0 |( {\bf \tilde{a}} \odot {\bf \tilde{a}})^2 \right\} +  a_{N^3LO}^{\mathrm{S} ,42} \, \left\{ ({\bf a}^\dagger \odot {\bf a}^\dagger )^2 | n^\prime \, 0 \rangle \langle n \, 0 | {\bf \tilde{a}} \odot {\bf \tilde{a}} + {\bf a}^\dagger \odot {\bf a}^\dagger | n^\prime \, 0 \rangle \langle n \, 0 |( {\bf \tilde{a}} \odot {\bf \tilde{a}})^2 \right\}  \nonumber \\
 &&~~~~~~~~~~~~~~~~+   a_{N^3LO}^{\mathrm{S} ,60} \, \left\{ ({\bf a}^\dagger \odot {\bf a}^\dagger )^3 | n^\prime \, 0 \rangle \langle n \, 0 | +  | n^\prime \, 0 \rangle \langle n \, 0 |( {\bf \tilde{a}} \odot {\bf \tilde{a}})^3 \right\} \Big] \nonumber \\
 && V_\delta^{\mathrm{SD}} = \sum_{n^\prime n} d_{n^\prime n}^{\,0 \,0} \, \Big[ a_{NLO}^{\mathrm{SD}} \left\{  \left[  {\bf a}^\dagger \otimes {\bf a}^\dagger \right]_2 \, | n^\prime \, 0 \rangle \langle n \, 0 |  + | n^\prime \, 0 \rangle \langle n \, 0 | \,  \left[  {\bf \tilde{a}} \otimes {\bf \tilde{a}} \right]_2 \right\} +o(\mathrm{NNLO}) +
 o(\mathrm{N}^3\mathrm{LO})  \Big] \odot \left[ \boldsymbol{\sigma}_1 \otimes \boldsymbol{\sigma}_2 \right]_2
 \end{eqnarray}
 where the low-energy constants (LECs) $a_{LO}, a_{NLO}, ...$ carry units of energy.  The HO matrix elements are
 \begin{eqnarray} 
 &&\langle n^\prime (\ell^\prime=0\,S)JM;T M_T  | V_\delta^\mathrm{S} | n(\ell=0\,S)JM;  T M_T \rangle =  d_{n^\prime n}^{\,0 \,0}~
\biggl[ a_{LO}^{\mathrm{S}} -2 \bigl[(n^\prime-1)+(n-1) \bigr] a_{NLO}^{\mathrm{S}} +4 (n^\prime-1)(n-1)a_{NNLO}^{\mathrm{S},22}   \nonumber \\
&&+\,4 ((n^\prime-1)(n^\prime-2)+(n-1)(n-2)) a_{NNLO}^{\mathrm{S},40} -8 ((n^\prime-1)(n^\prime-2)(n-1)+ 
(n^\prime-1)(n-1)(n-2))a_{N3LO}^{\mathrm{S},42} \nonumber \\ 
&& \hspace{0.75in} +\,8((n^\prime-1)(n^\prime-2)(n^\prime-3)+(n-1)(n-2)(n-3))a_{N3LO}^{\mathrm{S},60} \biggr] \nonumber  \\
&&\langle n^\prime (\ell^\prime=0\,S=1)J=1M; TM_T | V_\delta^{\mathrm{SD}} | n(\ell=2\,S=1)J=1M;TM_T\rangle = 
 {4 \sqrt{2} \over 3}~ d_{n^\prime n}^{\,0 \, 2}~
\biggl[ a_{NLO}^\mathrm{SD} -2\bigl[(n^\prime-1) a_{NNLO}^{\mathrm{SD},22} \nonumber \\
&&\hspace{0.3in}+ (n-1)a_{NNLO}^{\mathrm{SD},04} \bigr] +4\bigl[(n^\prime-1)(n^\prime-2)a_{N^3LO}^{\mathrm{SD},42} + 
(n^\prime-1)(n-1) a_{N^3LO}^{\mathrm{SD},24} +
(n-1)(n-2)a_{N^3LO}^{\mathrm{SD},06} \bigr] \biggr] 
\label{eq:deltaVMe}
\end{eqnarray}
 \end{widetext}
HOBET's short-range expansion in oscillator quanta is equivalent to a gradient expansion around $r \sim b$, 
producing a characteristic dependence on nodal quantum numbers \cite{WH}.
 Similar expressions exist at N$^3$LO for the D, DG, P, PF, and F channels \cite{WK}.
 By exploiting properties of the free Green's function of Eq. (\ref{eq:freeG}), the corresponding
 ``edge states" matrix elements can also be evaluated analytically. Edge states are generating from $|n \ell \rangle$ in the last shell of $P$ where
 $EG_{QT}(E)P|n \ell \rangle \ne P|n \ell \rangle$: the kinetic energy IR sum in $Q$ is incorporated through these states.

The analogous short-range expansion for potentials is in terms 
of the Talmi moment integrals, e.g., for S-waves,
 \[ \int d \boldsymbol{r}^\prime d \boldsymbol{r}~ r^{2 p^\prime} e^{-{r^\prime}^2/2} Y_{00}(\Omega^\prime)\,
 V(\boldsymbol{r^\prime},\boldsymbol{r}) \, r^{2 p} e^{-{r}^2/2} Y_{00}(\Omega) \]
The LECs for a potential are proportional to these integrals.  Consequently the short-range contributions of $V_\pi$, denoted $V_\pi^{UV}$,
 can be exactly encoded in LECs: the expressions are given in \cite{WH}.  In pionful HOBET,
 $V_\pi^{IR} \equiv V_\pi -V_\pi^{UV}$.   Even before this subtraction,
 $V_\pi$, with its $1/r^3$ tensor force, is well-behaved in HOBET because $b$ and $\Lambda$ act
 as cutoffs.  Then HOBET executes additional subtractions up to the number of LECs available
 in each partial-wave channel, to produce a very soft $V_\pi^{IR}$.  
 
 When HOBET's LECs are determined, the pion makes no explicit contribution to any
 of the fitted channels, as $V_\pi^{UV}$ has been removed.  This contrasts with EFT approaches,
 where the pion is an explicit degree of freedom, employed at short ranges
 where realistic interactions are not pionic, generating
 debates about the best power-counting scheme \cite{Weinberg}.   Taking N$^3$LO as an example, once 
 HOBET's LECs are fixed, the values of the shortest range matrix elements
 for which $2(n^\prime-1)+\ell^\prime + 2(n-1)+\ell \le  6$ are completely determined.  The fitted LECs also contribute to 
 matrix elements for which $2(n^\prime-1)+\ell^\prime + 2(n-1)+\ell > 6$, but there 
 higher-order, long-range Talmi integrals also contribute.  Pionful HOBET assigns to these Talmi integrals  their
 pionic values.

 HOBET's summation of the effective kinetic energy operator to all orders in Eq. (\ref{eq:BH}) can generate differential
 reductions in
 binding energies of up to $\hbar \omega$, as $E \rightarrow 0$ from below,
 illustrating the size of the shifts associated with kinetic energy delocalization in this limit.
 The edge state can be computed from the free Green's function, at the cost of a matrix inversion in $P$
 \begin{eqnarray}
 E G_{QT}P |n \ell m \rangle &=& G_0(E) [P G_0(E) P]^{-1} |n \ell m \rangle \nonumber \\
 G_0(E) &=& \left\{ \begin{array}{ll}  1/(\boldsymbol{\nabla}^2 - \kappa^2) & E<0 \\ 1/(\boldsymbol{\nabla}^2 + k^2) & E>0 \end{array} \right.
 \label{eq:freeG}
 \end{eqnarray}
where $\kappa \equiv \sqrt{2 |E|/\hbar \omega}$, $k=\sqrt{2E/\hbar \omega}$, and $\boldsymbol{\nabla}$ are
dimensionless.  Matrix elements of $G_0$ in P can be evaluated analytically.  
We employ standing-wave Green's functions.  The proper treatment of these Green's functions
is the key to executing HOBET as an ET.

For bound states $G_0(E)$ is determined by the binding energy, with self-consistent solutions of the BH
equation then yielding the discrete eigenvalues.  In contrast, for $E>0$, there exists a solution at every energy,
while the IR behavior depends not only on $E$, but also on the phase shift
$\delta_\ell(E)$ that appears in the homogeneous term in $G_0$,
\begin{eqnarray}
&&G_0(E>0;{\bf r}, {\bf r}^\prime) =\displaystyle{ - {\cos{k|{\bf r} - {\bf r}^\prime|} \over 4 \pi |{\bf r}-{\bf r}^\prime|} } \nonumber \\
&&~~- k \sum_{\ell m} \cot{\delta_\ell(E)}~ j_\ell(k { r}) ~j_\ell(k { r}^\prime)~Y_{\ell m}(\Omega) Y_{\ell m}^*(\Omega^\prime) 
\end{eqnarray}
Phase shifts previously encoded into NN potentials are thus fed into HOBET through $G_0$, properly fixing its IR
behavior.  If this is done at some $E$ followed by a diagonalization in $P$, clearly an eigenvalue 
will not typically be found at that
$E$.  As the theory is complete and the IR behavior correct, the source of this discrepancy must
be in the UV, an inadequate $V_\delta$.  $V_\delta$'s  LECs should then be adjusted to fix the discrepancy.
In practice the N$^3$LO LECs are chosen to produce a best fit to all of the phase shift information between threshold 
and 40 MeV CM energy, in the $^1$S$_0$, $^3$S$_1$-$^3$D$_1$, $^1$D$_2$, $^3$D$_1$, $^3$D$_2$, $^3$D$_3$-$^3$G$_3$,
$^1$P$_1$, $^3$P$_0$, $^3$P$_1$, $^3$P$_2$-$^3$F$_2$, $^1$F$_3$, and $^3$F$_\mathrm{J}$ channels, in channel by channel calculations.
The number of LECs at N$^3$LO varys from six in the S-wave channels to one in the F and mixed DG channels.
Technical details of the fitting procedure are given in \cite{WK}.  
\iffalse
Singlet channels are
simpler than triplet channels, where tensor forces mix states of $\ell$ and $\ell+2$, and thus where the number 
of relevant LECs is roughly double.  For these cases, it is helpful numerically to work in a mixed basis of,
say, S and D waves, corresponding to a diagonalized S-matrix.
\else
When tensor forces mix channels such as $^3$S$_1/^3$D$_1$, we use the S-matrix to express the possible 
asymptotic states as a linear combination of basis states.   We construct an $H^{eff}$ for each basis state to
constrain the entire space.  A convenient basis is implied by the diagonalization of the S matrix, giving one asymptotic state that
is mostly $^3$S$_1$ and one that is mostly $^3$D$_1$.    We can assign a number $n_b$ of weights, $W_b$, to these states reflecting the fraction of
 $^3$S$_1$ and  $^3$D$_1$ in the bound state.
\fi

With an exact ET and perfect phase shift data, $PH^\mathrm{eff}(E_i) P|\Psi_i \rangle = E_i P |\Psi_i \rangle$
for each energy $E_i$ of a set spanning the energy interval of interest.  But
under realistic conditions,
$PH^\mathrm{eff}(E_i) P|\Psi_i \rangle = \epsilon_i P |\Psi_i \rangle$, where $\epsilon_i$ is an eigenvalue
near but not identical to $E_i$.  Our LECs were determined by 
minimizing the total self-consistency error over sampled energies, 
C$^2$(LECs)=$\left(1/(N*n_b)\right)\sum_{b=1}^{n_b}\sum_{i=1}^N$ C$^2_{b,i}$ where C$_{b,i}=W_b (E_i-\epsilon_i)/E_i \equiv W_b \Delta E_i/E_i$.
The fitting was done successively through the orders, e.g., with the LECs from the NLO minimization
forming the starting values for the NNLO LEC minimization, etc.

\begin{figure}[ht]   
\centering
\includegraphics[scale=0.335 ]{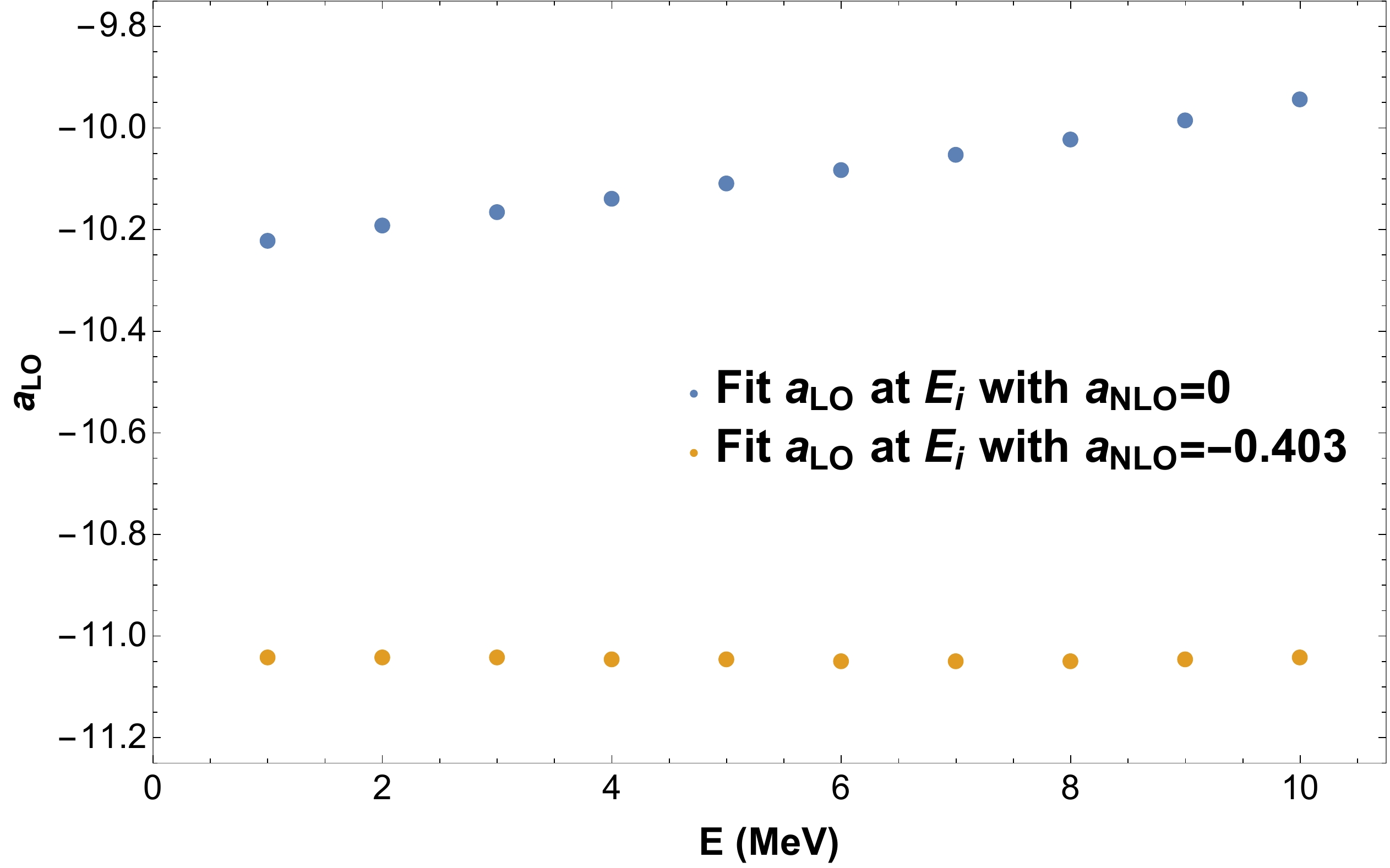}
\caption{Energy dependence of $a_{LO}$ at LO (upper dots) and residual energy dependence $a_{LO}$ at NLO (lower dots) after $a_{NLO}$ is fixed at -0.403.     }
\label{Fig:LECEnergyDep}
\end{figure}

The procedure was tested in a realistic S-wave model -- square well plus hard core --  for which exact phase shifts were known,
in order to evaluate convergence properties.   A $P$ space with $b=1.7$ fm and $\Lambda=8$ (5 included S-states) was used in this test.  A $\sim$ two-orders-of-magnitude improvement in $C^2$ per order in the expansion \cite{WH} was obtained.



Another important test was performed with the same model and P space: are LECs obtained in one energy interval
indeed constant, over the full energy range considered?   $a_{LO}$ was determined 10 times at single energies
sampled at 1 MeV intervals from 1 to 10 MeV, with higher order LECs set to zero.  The results in Fig. \ref{Fig:LECEnergyDep}
show a residual energy dependence in $a_{LO}$ of about 3\% over the range.  A second fit was then done at NLO using
two sample energies, 1 and 10 MeV, to determine $a_{NLO}$.  Keeping $a_{NLO}$ fixed, the first step,
determining $a_{LO}$ at each of the 10 sample energies, was repeated.  Fig. \ref{Fig:LECEnergyDep}
shows almost no energy dependence in the new $a_{LO}$s ($< 0.1$\%).   
This behavior is a general property of our fits and highly desirable in an ET:
residual energy dependence of the LECs simply reflects corrections from orders beyond the
last included order, the largest contribution coming from the immediately following order.
 
We then applied the method to experimental NN data \cite{av18a}, fitting 
phase shifts at 40 energies uniformly spaced from 1 to 40 MeV, again defining the P space with $\Lambda=8$  and $b=1.7$ fm.
For the  $^3$S$_1/^3$D$_1$ channel we used weights $W_S=1$ and $W_D=0.1$
for the asymptotic basis states corresponding to the diagonalized S-matrix.    
From the resulting LECs we calculated the deuteron binding energy.
The results are shown in Table \ref{tab:deuteron}, for
pionless and pionful HOBET, as a function or order, including the self-consistency error.
While both calculations converge well,
the comparison shows the importance of the chiral IR correction in pionful HOBET:
at N$^3$LO the deuteron binding energy is correct to 0.1 keV, and the phase-shift fit (reflected in
the self-consistency error) is nearly perfect.
 \iftrue
 \begin{table}[b]
\caption{\label{tab:SPotEnergyTable}  Deuteron channel: binding energy $E_b$ as a function of the expansion
order.  Bare denotes a calculation with $T+V$ .}
\begin{ruledtabular}
\begin{tabular}{lcccc}
\textrm{Order}&
\textrm{$E_\mathrm{b}^\mathrm{pionless}$}&
\textrm{$C^2\left(LECs\right)$} & \textrm{$E_\mathrm{b}^\mathrm{pionful}$}&
\textrm{$C^2\left(LECs\right)$} \\
\colrule
bare & 3.09525 & - &   -0.76775 & - \\
LO      &  -1.27715 & 2.2E-2 &     -2.01110 & 1.9E-3 \\
NLO   &  -1.95424 & 1.6E-2  &    -2.19833 & 2.2E-6 \\
NNLO & -2.17307 & 6.7E-3 &     -2.21705 & 4.0E-8  \\
N$^3$LO & -2.23175 & 1.3E-3 & -2.22464 & 8.4E-9 
\end{tabular}
\end{ruledtabular}
\label{tab:deuteron}
\end{table}
\else
 \begin{table}
\caption{\label{tab:SPotEnergyTable}  Deuteron channel: binding energy $E_b$ as a function of the expansion
order.  Bare denotes a calculation with $T+V$ while 0 denotes one with the IR correction,
but $V_\delta=0$.}
\begin{ruledtabular}
\begin{tabular}{lcccc}
\textrm{Order}&
\textrm{$E_\mathrm{b}^\mathrm{pionless}$}&
\textrm{$C^2\left(LECs\right)$} & \textrm{$E_\mathrm{b}^\mathrm{pionful}$}&
\textrm{$C^2\left(LECs\right)$} \\
\colrule
bare & 3.09525 & - & -0.76775 & - \\
0 (IR) & unbound & 7.3E-2  & -0.13067 & 5.4E-2 \\
LO      &  -1.27715 & 2.2E-2 & -2.01412 & 2.0E-3 \\
NLO   &  -1.95461& 1.6E-2  & -2.16739 & 7.4E-6 \\
NNLO & -2.18608 & 6.7E-3 & -2.21724 & 1.5E-6  \\
N$^3$LO & -2.21644 & 1.3E-3 & -2.22222 & 9.9E-7 
\end{tabular}
\end{ruledtabular}
\label{tab:deuteron}
\end{table}
\fi

\begin{figure}[tp]   
\centering
\vspace{2mm}
\includegraphics[scale=0.255 ]{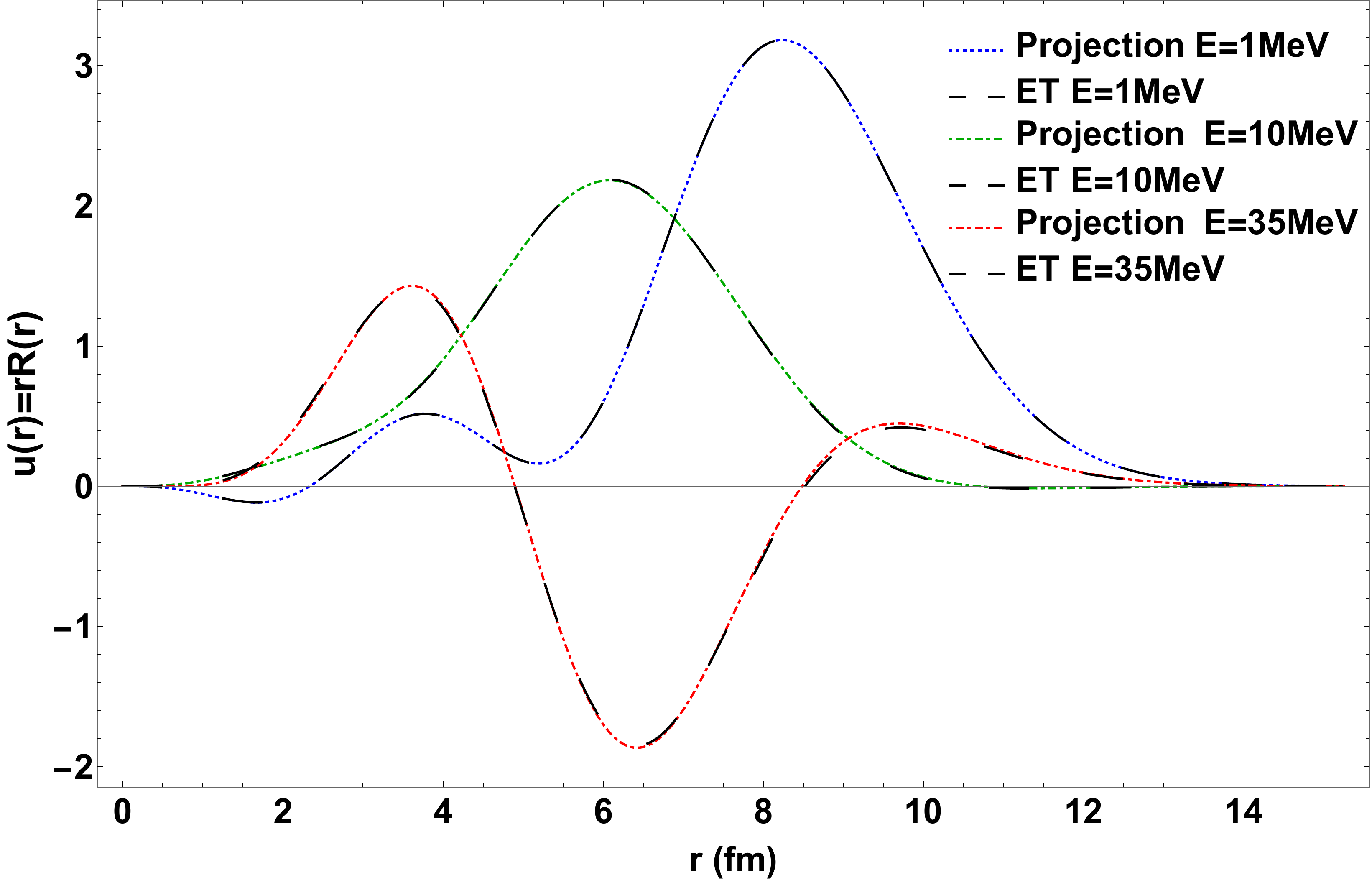}
\caption{Projections of exactly computed $^1P_1$ relative wave functions are shown to match the HOBET wave
 functions nearly perfectly, for representative continuum energies.}
\label{Fig:Fit1P1}
\end{figure}

An additional test is the quality of the projected
wave function.   Fig. \ref{Fig:Fit1P1} shows that the $^1$P$_1$ results at CM energies 1, 15, and 35 MeV are in
nearly perfect agreement with the exact results: all of the detailed behavior of the projected wave functions as continuous functions of $r$ and $E$, remarkably, can be encoded in a few LECs.  The S-wave LECs obtained in
at N$^3$LO are given in Table \ref{table:1}; results for other channels are given in \cite{WK}. 

\begin{table}
\caption{The S-wave LECs determined at N$^3$LO in pionless and pionful HOBET.  See
\cite{WK} for the full set of couplings.}
\begin{ruledtabular}
\label{table:1}
\begin{tabular}{cccc}
Transitions & LECs (MeV) & Pionless& Pionful \\
 \colrule
${}^3S_1 \leftrightarrow {}^3S_1$  & $a_{LO}^{3S1} $ & -49.9309 & -54.8429  \\
                                                      & $a_{NLO}^{3S1}$  & -5.70068 & -8.16310 \\
                                                      & $a_{NNLO}^{3S1,22}$ &  -9.73003E-1 & -2.07700  \\
                                                      & $a_{NNLO}^{3S1,40}$ &  -1.93934E-1 & -2.4235E-1  \\
                                                      & $a_{N3LO}^{3S1,42}$  & -5.61191E-2 & -2.3738E-1  \\
                                                      & $a_{N3LO}^{3S1,60}$  & -8.70527E-2 & 4.3667E-4  \\
\rule{0pt}{3ex} ${}^1S_0 \leftrightarrow {}^1S_0$  & $a_{LO}^{1S0} $ &  -38.5110 &  -39.2041  \\
                                                      & $a_{NLO}^{1S0}$ & -9.40213 & -6.88560  \\
                                                      & $a_{NNLO}^{1S0,22}$ & -4.23143 & -1.90118  \\
                                                      & $a_{NNLO}^{1S0,40}$ & 1.27787E-1 & -3.75499E-1  \\
                                                      & $a_{N3LO}^{1S0,42}$ & -4.51098E-1 & -2.45101E-1  \\
                                                      & $a_{N3LO}^{1S0,60}$ & -2.02571E-1 & -3.63233E-3  
\end{tabular}
\end{ruledtabular}
\end{table}
In summary, we have demonstrated a short-range ET expansion for structure and reactions that is 
convergent and continuous in $E$, directly linking experimental scattering observables to
bound state properties.  Phase shifts enter through a BH equation reorganization that builds in
correct IR behavior.   This procedure simplifies subsequent calculations in larger
nuclei, reducing the A-body problem to one involving iterated softened strong interactions described 
by finite matrices
\[  V \rightarrow P  E G_{TQ}(E)  \left[  V_\pi^{IR} + V_\delta \right] E G_{QT}(E)  P \]
separated by IR propagation in $Q$.
An exact treatment of  the two-body physics of such systems by summing this series is the next step in HOBET development.

{\it Acknowledgement:} This material is based upon work supported in part by the US DOE, Office of Science,
Office of Nuclear Physics and SciDAC under awards DE-SC00046548,
DE-AC02-05CH11231, and KB0301052.


\begin{thebibliography}{000}
\bibitem{lattice} Finite box boundary conditions can also be adopted, for matching to lattice QCD eigenvalues.
\bibitem{av18a} To facilitate comparisons, we used as ``experimental" phase shifts those implicit in Argonne $v18$,
which was fit to experimental data.
\end{thebibliography}
\end{document}